\documentclass[prd, amsfonts, twocolumn, nofootinbib, showpacs]{revtex4}
\usepackage{graphicx}
\usepackage{epsfig}
\usepackage{color}
\usepackage{amsmath}
\usepackage{amssymb}
\newcommand{\be}{\begin{equation}}
\newcommand{\ee}{\end{equation}}
\newcommand{\bea}{\begin{eqnarray}}
\newcommand{\eea}{\end{eqnarray}}

\newcommand{\gapp}{\mathrel{\raise.3ex\hbox{$>$}\mkern-14mu
\lower0.6ex\hbox{$\sim$}}}
\newcommand{\lapp}{\mathrel{\raise.3ex\hbox{$<$}\mkern-14mu
\lower0.6ex\hbox{$\sim$}}}
\def\bbox{{\,\lower0.9pt\vbox{\hrule \hbox{\vrule height 0.2 cm
\hskip 0.2 cm \vrule  height 0.2 cm}\hrule}\,}}

\begin{document}
\title{Response to the Comment on ``Observing a wormhole" }
\author{De-Chang Dai\footnote{corresponding author: De-Chang Dai$^{1,2}$,\\ email: diedachung@gmail.com\label{fnlabel}}, Dejan Stojkovic$^3$,}
\affiliation{$^1$ Center for Gravity and Cosmology, School of Physics Science and Technology, Yangzhou University, 180 Siwangting Road, Yangzhou City, Jiangsu Province, P.R. China 225002 }
\affiliation{ $^2$ Department of Physics, Case Western Reserve University,
10900 Euclid Avenue, Cleveland, OH 44106 }
\affiliation{ $^3$ HEPCOS, Department of Physics, SUNY at Buffalo, Buffalo, NY 14260-1500, U.S.A.}


\begin{abstract}
\widetext
In the Comment written by Krasnikov \cite{Krasnikov:2019mrd} on our recent paper \cite{Dai:2019mse}, the author argues that a spherically symmetric perturbation coming from a massive object located on the other side of the wormhole would just add up to the total mass of the central object, and is therefore useless as an indicator of the wormhole presence.   
We point out that our Eq.~(37) represents an {\it acceleration variation} in the motion of the star S2 due to an elliptic (and thus non-spherically symmetric) orbit of the star on the other side perturbing the metric. This time-dependent acceleration variation is in principle distinguishable from the original acceleration coming from the central object. We also point out that the author is trying to apply Birkhoff's
theorem in a setup where it cannot be applied in a straightforward way. 
\end{abstract}

\maketitle

We would like to thank the author for finding our paper interesting. We also appreciate his efforts to clarify some issues with our calculations. However, it is easy to see that all the points that the author raises are not correct.

The author's comment \cite{Krasnikov:2019mrd} is based on observation that the whole setup is spherically symmetric, which would by Birkhoff' theorem imply that the extra force that comes from an object located on the ``other side" just adds up to the original central force of the super-massive object, and is indistinguishable from it.

In the language of symmetries, we do assume a spherically symmetric background, but the perturbations coming from the elliptic orbit violates this symmetry. Birkhoff’s theorem cannot be applied to our perturbed universe anymore.

The author's confusion perhaps stems from the fact that our  Eq.~(36) is a result of a monopole perturbation. However, in section V of our paper \cite{Dai:2019mse}, we clearly say that we consider an elliptic orbit of a star located on the ``other side". An elliptic orbit cannot be represented with only one monopole, and can be viewed instead as a sequence of monopoles. We estimate the effect in our Eq.~(37) by using two monopoles, one for a perigee, $r_p$, and another for an apogee, $r_a$, as 
\begin{equation}
\Delta a =\mu \left(\frac{R}{r_p}-\frac{R}{r_a}\right)\frac{1}{r_2^2} .
\end{equation}
where $\Delta a$ is the acceleration variation, $\mu$ is the mass of the star perturbing the metric, $R$ is size of the wormhole mouth, and $r_2$ is the radial coordinate in our space. Thus, our Eq.~(37) represents an {\it acceleration variation} in the motion of the star S2 due to an elliptic orbit of the star on the other side perturbing the metric. This time-dependent acceleration variation is in principle distinguishable from the original acceleration coming from the central object.

It is known that propagation of a source-free gravitational signal (emitted by local disturbance that quickly disappears) requires a non-vanishing quadrupole. However, the cause of S2 star’s orbit deviation is a continuous change of the Newtonian force between the source of perturbations and S2 star. The force that S2 star feels depends on the distance to the perturber, and this distance changes with a certain period. Monopole term determines only the change in the total force. We used two monopoles (one for a perigee and another for an apogee) to estimate the magnitude of the time-dependent variation in Newtonian force.  Thus, S2's orbit is perturbed by time-varying force, not by gravitational waves.

In his {\it Remark}, the author of the Comment makes even a stronger statement. He claims not only that our anomalous acceleration should be indistinguishable from the acceleration coming from the central object, but it should actually be zero, again because of Birkhoff' theorem. To justify his claim, he quotes \cite{Chen:2016plo}: “Inside the orbit, the perturbation vanishes.”

Our expression differs from that in \cite{Chen:2016plo} in the second term on the right hand side of
Eq.~(28). The reason is that \cite{Chen:2016plo} uses Zerilli gauge, which is not appropriate in our case. We require 
that $g_{tt}$ is continuous at the shell of the radius
$R$ (the radius of the wormhole mouth). Our physical requirement is that the flux is continuous at the connection of the two space-times, so we have a continuous metric (and its derivatives) there.  Continuous $g_{tt}$ and $g_{rr}$ can be found in the Lorentz gauge (see for example Eqs.~(46)-(51) and Fig.~1 in \cite{Barack:2005nr}). Therefore, in our case perturbations do not vanish inside the orbit.

The author's confusion in his {\it Remark } comes from naive application of Birkhoff's theorem. The main point is that Birkhoff’s theorem is of local nature and
can work in a certain region of space-time (for example outside
a ``breathing'' spherically symmetric star), therefore any criticism
mentioning its application to the whole space-time is nor warranted. 

More specifically, in the context of our model, a wormhole connects two initially disconnected spaces, and Birkhoff's theorem cannot be applied in a straightforward way. One could see this by going through the derivation of the Schwarzschild solution. In order to make a general spherically symmetric space-time static, one has to redefine the time coordinate in order to absorb the time dependence. But in the case of a wormhole this freedom is limited by the requirement that two space-times are smoothly connected at the wormhole mouth. Thus, trivial application of the Birkhoff's theorem is not possible.  Another way to see the same thing is to consider an empty spherically symmetric shell on one side of the wormhole. By Birkhoff's theorem, the force inside the shell is zero. But the very presence of the massive shell modifies the asymptotics on one side of the wormhole, and thus violates the symmetry properties of the original wormhole solution (two identical space-times connected by a throat). 

Finally, in the footnote [5] of his Comment, the author states that we did not justify our matching conditions at the wormhole mouth $r=R$. In the thin-shell and short-throat
wormhole approximation that we are working these matching conditions are actually natural. The acceleration must be continuous, if the source of perturbations is not on the shell (wormhole mouth), see for example Eqs.~21.175b and 21.175c in \cite{MTW}. See also the same choice of conditions (above Eq.~(4)) in \cite{Cardoso:2016rao}.

We would therefore like to thank the author of the comment for his interest in our work, however we have to say that it is easy to see that his objections are unfounded.

\end{document}